\documentclass[lineno]{biometrika}

\usepackage{amsmath}

\usepackage{times}
\usepackage{bm}
\usepackage{natbib}

\usepackage{epstopdf}
\usepackage[plain,noend]{algorithm2e}

\makeatletter
\renewcommand{\algocf@captiontext}[2]{#1\algocf@typo. \AlCapFnt{}#2} 
\renewcommand{\AlTitleFnt}[1]{#1\unskip}
\def\@algocf@capt@plain{top}
\renewcommand{\algocf@makecaption}[2]{%
  \addtolength{\hsize}{\algomargin}%
  \sbox\@tempboxa{\algocf@captiontext{#1}{#2}}%
  \ifdim\wd\@tempboxa >\hsize
    \hskip .5\algomargin%
    \parbox[t]{\hsize}{\algocf@captiontext{#1}{#2}}
  \else%
    \global\@minipagefalse%
    \hbox to\hsize{\box\@tempboxa}
  \fi%
  \addtolength{\hsize}{-\algomargin}%
}
\makeatother

\def\Bka{{\it Biometrika}}

\def\v{{\varepsilon}}

\def\AR{\textsc{AR}}

\begin{document}

\jname{}
\jyear{}
\jvol{}
\jnum{}
\copyrightinfo{}


\markboth{H. Kang, T. T. Cai \and D. S. Small}{A simple and robust confidence interval for causal effects with possibly invalid instruments}

\title{A simple and robust confidence interval for causal effects with possibly invalid instruments}

\author{H. KANG, T. T. CAI, \and D. S. SMALL}
\affil{Department of Statistics, University of Pennsylvania, Philadelphia, PA 19104, U.S.A. \email{khyuns@wharton.upenn.edu} \email{tcai@wharton.upenn.edu} \email{dsmall@wharton.upenn.edu}}

\maketitle

\begin{abstract}
Instrumental variables have been widely used to estimate the causal effect of a treatment on an outcome. Existing confidence intervals for causal effects based on instrumental variables assume that all of the putative instrumental variables are valid; a valid instrumental variable is a variable that affects the outcome only by affecting the treatment and is not related to unmeasured confounders. However, in practice, some of the putative instrumental variables are likely to be invalid. This paper presents a simple and general approach to construct a confidence interval that is robust to possibly invalid instruments. The robust confidence interval has theoretical guarantees on having the correct coverage and can also be used to assess the sensitivity of inference when instrumental variables assumptions are violated. The paper also shows that the robust confidence interval outperforms traditional confidence intervals popular in instrumental variables literature when invalid instruments are present. The new approach is applied to a developmental economics study of the causal effect of income on food expenditures.
\end{abstract}


\section{Introduction} \label{sec:intro}

The instrumental variables method is a popular method to estimate the causal effect of a treatment, exposure, or policy on an outcome when unmeasured confounding is present \citep{angrist_identification_1996, tan_regression_2006, baiocchi_instrumental_2014}. Informally speaking, the method relies on having instruments that are (A1) related to the exposure, (A2) only affect the outcome by affecting the exposure (no direct effect), and (A3) are not related to unmeasured confounders that affect the exposure and the outcome (see Section \ref{sec:model} for details). Unfortunately, in many applications, practitioners are unsure if all of the candidate instruments satisfy these assumptions. For example, in Mendelian randomization, the candidate instruments are genetic variants which are associated with the exposure and may also have a direct effect on the outcome, an effect known as pleiotropy, thereby violating (A2), or may be in linkage disequilibrium, thereby violating (A3) \citep{davey_smith_mendelian_2003, davey_smith_mendelian_2004, lawlor_mendelian_2008,burgess_use_2012,solovieff_pleiotropy_2013}. A similar problem arises in economics where some instruments may violate the exogeneity assumption, which is a combination of (A2) and (A3) \citep{murray_avoiding_2006,conley_plausibly_2012}.


Violation of (A1), known as the weak instrument problem, has been studied in detail; see \citet{stock_survey_2002} for a survey. In contrast, the literature on violations of (A2) and (A3), known as the invalid instrument problem \citep{murray_avoiding_2006}, is limited. \citet{andrews_consistent_1999} and \citet{andrews_consistent_2001} considered selecting valid instruments within the context of generalized method of moments \citep{hansen_large_1982}, but not inference of the treatment effect after selection of valid instruments. \citet{liao_adaptive_2013} and \citet{cheng_select_2015} considered estimation of the treatment effect when there is, a priori, a known, specified set of valid instruments and another set of instruments which may not be valid. Our work is most closely related to the recent works by \citet{kolesar_identification_2013, bowden_mendelian_2015} and \citet{kang_instrumental_2015}. \citet{kolesar_identification_2013} and \citet{bowden_mendelian_2015} considered the case when the instruments violate (A2) and proposed an orthogonality condition where the instruments' effect on the exposure are orthogonal to their effects on the outcome. \citet{kang_instrumental_2015} considered more general violations of (A2) and (A3) based on imposing an upper bound on the number of invalid instruments among the candidate instruments, without knowing which instruments are valid a priori or without imposing  structure on the instruments's effect like \citet{kolesar_identification_2013} and \citet{bowden_mendelian_2015}. However, \citet{kang_instrumental_2015} only studied point estimation and not confidence intervals.

This paper focuses on developing confidence intervals when candidate instruments may be invalid, within the framework introduced by \citet{kang_instrumental_2015}. We propose a simple and general confidence interval procedure that theoretically guarantees the correct coverage rate in the presence of invalid instruments and can easily be used with traditional instrumental variables methods. Our robust confidence interval can also be interpreted within the context of a sensitivity analysis in causal inference where the interval measures the change in inference about the treatment effect if instruments violate (A2) and (A3). We also provide a new theoretical framework to study confidence intervals properties, including size and power, under invalid instruments. We also demonstrate that our method can produce valid and informative confidence intervals in both synthetic and real data.

\section{Setup}

\subsection{Notation}

We use the potential outcomes notation \citep{rubin_estimating_1974} for instruments laid out in \citet{holland_causal_1988}. Specifically, let there be $L$ potential candidate instruments and $n$ individuals in the sample. Let $Y_i^{(d,z)}$ be the potential outcome if individual $i$ had exposure $d$, a scalar value, and instruments $z$, an $L$ dimensional vector. Let $D_i^{(z)}$ be the potential exposure if the individual had instruments $z$. For each individual, we observe the outcome $Y_i$, the exposure, $D_i$, and instruments $Z_{i}$. In total, we have $n$ observations of $(Y_i, D_i,Z_{i})$. We denote $Y = (Y_1,\ldots, Y_n)$, $D = (D_1,\ldots,D_n)$ and $Z$ to be the $n$ by $L$ matrix where row $i$ consists of $Z_{i}^T$ where $Z$ is assumed to have full rank.

For a subset $A \subseteq \{1,\ldots,L\}$, denote its cardinality $c(A)$ and $A^C$ its complement. Also, let $Z_{A}$ be an $n$ by $c(A)$ matrix of instruments where the columns of $Z_A$ are from the set $A$, $P_{Z_A} = Z_A (Z_A^T Z_A)^{-1}Z_{A}^T$ be the orthogonal projection matrix onto the column space of $Z_{A}$ and $R_{Z_A}$ be the residual projection matrix so that $R_{Z_A} + P_{Z_A} = I$ where $I$ is an $n$ by $n$ identity matrix. Also, for any $L$ dimensional vector $\pi$, let $\pi_A$ only consist of elements of the vector $\pi$ determined by the set $A \subseteq \{1,\ldots,L\}$.

\subsection{Model and definition of valid instruments} \label{sec:model}

For two possible values of the exposure $d',d$ and instruments $z',z$, we assume the following potential outcomes model
\begin{equation} \label{eq:model1}
Y_{i}^{(d',z')} - Y_{i}^{(d,z)} =  (z' - z)^T \phi^* + (d' - d) \beta^*,\quad E\{Y_i^{(0,0)} \mid Z_{i}\} =  Z_{i}^T \psi^* 
\end{equation}
where $\phi^*, \psi^*$, and $\beta^*$ are unknown parameters. The parameter $\beta^*$ represents the causal parameter of interest, the causal effect (divided by $d' - d$) of changing the exposure from $d'$ to $d$ on the outcome. The parameter $\phi^*$ represents violation of (A2), the direct effect of the instruments on the outcome. If (A2) holds, then $\phi^* = 0$. The parameter $\psi^*$ represents violation of (A3), the presence of unmeasured confounding between the instrument and the outcome. If (A3) holds, then $\psi^* = 0$. 

Let $\pi^* = \phi^* + \psi^*$ and $\epsilon_i = Y_{i}^{(0,0)} - E\{Y_{i}^{(0,0)} \mid Z_{i}\}$. When we combine equations \eqref{eq:model1} along with the definition of $\epsilon_i$, the observed data model becomes
\begin{equation} \label{eq:model2}
Y_i =  Z_{i}^T \pi^* +  D_i \beta^* + \epsilon_i,\quad{} E(\epsilon_i \mid Z_{i}) = 0
\end{equation}
The observed model is also known as the under-identified single-equation linear model in econometrics (page 83 of \citet{wooldridge_econometrics_2010}). The model \eqref{eq:model2} can be generalized to include heterogeneous causal effects and non-linear effects; see \citet{kang_instrumental_2015} for details. Also, the model can incorporate exogenous covariates, say $X_{i}$, including an intercept term, and the Frisch-Waugh-Lovell Theorem allows us to reduce the model with covariates to \eqref{eq:model2} \citep{davidson_estimation_1993}. The parameter $\pi^*$ in the observed data model \eqref{eq:model2} combines both the violation of (A2), represented by $\phi^*$, and the violation of (A3), represented by $\psi^*$. If both (A2) and (A3) are satisfied, then $\phi^* = \psi^* = 0$ and $\pi^* = 0$. Hence, the value of $\pi^*$ captures whether instruments are valid versus invalid. Definition \ref{def:validIV} formalizes this idea.

\begin{definition} \label{def:validIV}
Suppose we have $L$ candidate instruments along with the models \eqref{eq:model1}--\eqref{eq:model2}. We say that instrument $j = 1,\ldots,L$ is valid if $\pi_j^* = 0$ and invalid if $\pi_j^* \neq 0$.
\end{definition}

When there is only one instrument, $L = 1$, Definition \ref{def:validIV} of a valid instrument is identical to the definition of a valid instrument in \citet{holland_causal_1988}. Specifically, assumption (A2), the exclusion restriction, which implies $Y_i^{(d,z)} = Y_i^{(d,z')}$ for all $d,z,z'$, is equivalent to $\phi^* = 0$ and assumption (A3), no unmeasured confounding, which means $Y_i^{(d,z)}$ and $D_{i}^{(z)}$ are independent of $Z_{i}$ for all $d$ and $z$, is equivalent to $\psi^* = 0$, implying $\pi^* = \phi^* + \psi^* = 0$. Definition \ref{def:validIV} is also a special case of the definition of a valid instrument in \citet{angrist_identification_1996} where here we assume the model is additive, linear, and has a constant treatment effect $\beta$. Hence, when multiple instruments, $L > 1$, are present, our models \eqref{eq:model1}--\eqref{eq:model2} and Definition \ref{def:validIV} can be viewed as a generalization of the definition of valid instruments in \citet{holland_causal_1988}. 



Let $s = 0,\ldots,L - 1$ be the number of invalid instruments and $U$ be an upper bound on $s$ plus $1$, i.e. the number of invalid instruments is assumed to be less than $U$. We assume that there is at least one valid IV, even if we don't know which among the $L$ IV is valid, since if all $L$ IVs are invalid (i.e. $s = L$), identification would not be possible \citep{kang_instrumental_2015}. 

This setup was also considered in \citet{kang_instrumental_2015} as a relaxation to traditional instrumental variables setups where one knows exactly which instruments are valid and invalid. Also, in Mendelian randomization where instruments are genetic, the setup represents a way for a genetic epidemiologist to impose prior beliefs about their genetic instruments' validity. For example, based on the investigator's expertise and the genome wide association studies, the investigator may provide an upper bound $U$, with smaller $U$s representing an investigator's confidence that most of their $L$ instruments are valid and vice versa.

Finally, the setup can be viewed as a sensitivity analysis commonly found in causal inference. In particular, similar to the sensitivity analysis presented in \citet{rosenbaum_observational_2002}, we can treat $U$ as the sensitivity parameter and vary from $U = 1$ to $U = L$ where $U = 1$ represents the traditional case where all instruments satisfy (A2) and (A3) and $U = L$ represents the worst case where at most $L - 1$ instruments may violate (A2) and (A3). For each $U$, we can construct confidence intervals for each $U$, and observe how violations of instrumental variables assumptions impact the resulting conclusions about the casual effect $\beta^*$. Similar to a typical sensitivity analysis, we can find $U$ where our confidence interval includes the null value of $\beta^* = 0$ and therefore invalidate the causal effect. If at $U = L$, the confidence interval does not contain the null value, then the conclusion about the causal effect $\beta^*$ is less sensitive to violations of the instrumental variables assumptions (A2) and (A3).

\section{Robust Confidence Intervals With Invalid Instruments}

\subsection{A general procedure} \label{sec:genprocedure}
Let $B^* \subset \{1,\ldots,L\}$ be the true set of invalid instruments. Then, $Z_{(B^*)^C}$ are the set of valid instruments and $Z_{B^*}$ are exogenous covariates we adjust for in model \eqref{eq:model2}. In the instrumental variables literature, there are many test statistics $T(\beta_0,B^*)$ of the null hypothesis $H_0: \beta^* = \beta_0$ versus the alternative $H_a: \beta^* \neq \beta_0$ where $B^*$ contains invalid instruments and $(B^*)^C$ contain valid instruments. Inverting the test $T(\beta_0,B^*)$ under size $\alpha$ provides a $1 - \alpha$ confidence interval for $\beta^*$, denoted as $C_{1 - \alpha}(Y,D,Z,B^*)$.
\begin{equation} \label{eq:confIntInversion}
C_{1 - \alpha}(Y,D,Z,B^*) = \{\beta_0 \mid T(\beta_0,B^*) \leq q_{1 - \alpha}\}
\end{equation}
where $q_{1 - \alpha}$ is the $1 - \alpha$ quantile of the null distribution of $T(\beta_0,B^*)$. These tests include the two-stage least squares, Anderson-Rubin test \citep{anderson_estimation_1949}, the conditional likelihood ratio test \citep{moreira_conditional_2003}, and many others; see Supplementary materials for details. 

Unfortunately, in our problem, we do not know the true set $B^*$ of invalid instruments, so we cannot directly use \eqref{eq:confIntInversion} to estimate confidence intervals for $\beta^*$. However, from Section \ref{sec:model}, we have a constraint on the number of invalid instruments, mainly $s < U$. We can use this constraint by taking unions of $C_{1 - \alpha}(Y,D,Z,B)$ over possible sets of invalid instruments $B \subset \{1,\ldots,L\}$ where $c(B) < U$. The confidence interval using the true set of invalid instruments $C(Y,D,Z,B^*)$ will be in this union since $c(B^*) < U$. Our proposal is exactly this, except that we restrict the subsets $B$ to be of size $c(B) = U - 1$.
\begin{equation} \label{eq:confIntProposal}
C_{1 -\alpha}(Y,D,Z) = \cup_{B,c(B) = U - 1} \{C_{1 - \alpha}(Y,D,Z,B) \} 
\end{equation}
Theorem \ref{theorem1} shows that the confidence interval in \eqref{eq:confIntProposal} has proper coverage in the presence of possibly invalid instruments.
\begin{theorem}
\label{theorem1}
Suppose model \eqref{eq:model2} holds and $s < U$. Given $\alpha$, consider any test statistic $T(\beta_0,B$) with the property that for any $B^* \subseteq B$, $T(\beta_0,B)$ has size at most $\alpha$ under the null hypothesis $H_0: \beta^* = \beta_0$. Then, $C_{1 - \alpha}(Y,D,Z)$ in \eqref{eq:confIntProposal} always has at least $1 - \alpha$ coverage even with invalid instruments.
\end{theorem}
The proof is in the appendix. The proposed confidence interval $C_{1 - \alpha}(Y,D,Z)$ is robust to invalid instruments. It is also simple and general. Specifically, for any test statistic $T(\beta_0,B)$ discussed above with a valid size for $B^* \subseteq B$, one simply takes unions of confidence intervals of $T(\beta_0,B)$ over subsets of instruments $B$ where $c(B) = U - 1$. In addition, a key feature of our procedure is that we do not have to iterate through all subsets of instruments where $c(B) < U$; we only have to examine the largest possible set of invalid instruments,  i.e. those subsets that are at the upper boundary of $U$, $c(B) = U - 1$, to guarantee $1 - \alpha$ coverage. 

A potential caveat to our procedure is computational feasibility. Even though we restrict the union to subsets of exactly size $c(B) = U - 1$, if there are many candidate instruments $L$ and $U$ is moderate large, $C_{1-\alpha}(Y,D,Z)$ becomes computationally burdensome. However, in many instrumental variables studies, it is difficult to find good candidate instruments and rarely does the number of these candidates instruments exceed $L = 20$, which modern computing can handle. Hence, our procedure in \eqref{eq:confIntProposal} is computationally tractable for most practical applications.

\subsection{Shorter interval with pretesting} \label{sec:pretest}
As shown in Theorem \ref{theorem1}, the overall interval $C_{1-\alpha}(Y,D,Z)$ which takes unions of subsets $B$ of confidence intervals $C_{1-\alpha}(Y,D,Z,B)$ has the desired coverage level as shown in Theorem \ref{theorem1}. Some subsets $B$ contain all the invalid instruments, leading to unbiased confidence intervals (i.e. contain true $\beta^*$ with probability greater than or equal to $1 -\alpha$), while others may not include all of the invalid instruments, leading to biased confidence intervals and more importantly, elongating $C_{1-\alpha}(Y,D,Z)$ to achieve the desired coverage level.  In order for $C_{1 -\alpha}(Y,D,Z)$ to have the desired coverage level, we just need the union to contain one unbiased confidence interval and ideally, we would like to remove these biased confidence intervals in the union of $C_{1-\alpha}(Y,D,Z)$ to end up with potentially shorter intervals while still maintaining the desired coverage level. In this section, we propose a way to do this by pretesting whether each of the subsets $B^C$ in the union of $C_{1-\alpha}(Y,D,Z)$ contain only valid instruments.

Specifically, for a $1 - \alpha$ confidence interval, consider the null hypothesis that $B^C$, for $c(B^C) \geq 2$, contains only valid instruments, $\pi_{B_C}^* = 0$. Suppose $S(B)$ is the corresponding test statistic for this null with level $\alpha_1 < \alpha$ and $q_{1 - \alpha_1}$ is the $1 - \alpha_1$ quantile of the null distribution of $S(B)$. Then, a $1 -\alpha$ confidence interval for $\beta^*$ that incorporates the pretest $S(B)$, denoted as $C_{1-\alpha}'(Y,D,Z)$ is
\begin{equation} \label{eq:confIntProposalScreen}
C_{1 - \alpha}'(Y,D,Z) = \cup_{B} \{C_{1-\alpha_2}(Y,D,Z,B) \mid c(B) = U - 1, S(B) \leq q_{1-\alpha_1} \}
\end{equation}
where $\alpha = \alpha_1 + \alpha_2$. For example, if the desired confidence level for $\beta^*$ is 95\% where $\alpha =  0\cdot 05$, we can set $\alpha_1 = 0.01$ and $\alpha_2 = 0.04$ where we would conduct the pretest $S(B)$ at $0.01$ level and obtain $C_{1-\alpha_2}(Y,D,Z,B)$ at the $0.04$ level. Theorem \ref{theorem2} shows that $C_{1 - \alpha}'(Y,D,Z)$ achieves the desired $1 - \alpha$ coverage of $\beta^*$ in the presence of possibly invalid instruments.
\begin{theorem} \label{theorem2} Suppose the assumptions in Theorem \ref{theorem1} hold. For any pretest $S(B)$ where $c(B^C) \geq 2$ and has the correct size under the null hypothesis that $B^C$ contains only valid instruments, $C_{1 - \alpha}'(Y,D,Z)$ always has at least $1 - \alpha$ coverage even with invalid instruments.  
\end{theorem}
Similar to Theorem \ref{theorem1}, the procedure in \eqref{eq:confIntProposalScreen} is general in the sense that any pretest $S(B)$ with the correct size guarantees that the confidence interval will have the desired level of coverage. For example, the Sargan test \citep{sargan_estimation_1958} can act as a pre-test for \eqref{eq:confIntProposalScreen}; see Supplementary Materials for details.

\subsection{Power under invalid instruments} \label{sec:power}
While many tests will satisfy the requirements for Theorems \ref{theorem1} and \ref{theorem2}, some tests will be better than others where ``better'' can be defined in terms of statistical power, length of the confidence interval, or coverage, all in the presence of invalid instrument. In this section, we provide one framework to study the power of common tests in instrumental variables literature under invalid instruments as they are applied to our robust confidence intervals. Section \ref{sec:simGeneral} studies properties of tests with respect to confidence interval coverage and length.

To setup this framework, we follow the weak instrument literature mentioned in Section \ref{sec:intro} where tests under violation of (A1) was studied in great detail \citep{staiger_instrumental_1997, andrews_optimal_2006, andrews_performance_2007}, except we modify it to study violations of (A2) and (A3). 
\begin{subequations} \label{eq:powerSetup}
\begin{align}
Y_i &= Z_{i}^T \pi^* + D_i \beta^* + \epsilon_i, \quad{} E(\epsilon_i, \xi_i | Z_i) =0 \\
D_i &= Z_{i}^T \gamma^* + \xi_i \\
\begin{pmatrix} \epsilon_i \\ \xi_i \end{pmatrix} &\sim N\left[\begin{pmatrix} 0 \\ 0 \end{pmatrix},\begin{pmatrix} \sigma_2^2  & \rho \sigma_1 \sigma_2 \\ \rho \sigma_1 \sigma_2  & \sigma_1^2 \end{pmatrix}\right]
\end{align}
\end{subequations}
The setup in \eqref{eq:powerSetup} is a special case of model \eqref{eq:model2} with the additional assumptions that (i) $D_i$ is related linearly to $Z_{i}$ and (ii) the error terms are bivariate i.i.d. Normal distributions with an unknown covariance matrix. Also, the two major differences between the setup considered in the weak instrumental variables work of \citet{staiger_instrumental_1997} and \citet{andrews_performance_2007} and the setup here is the introduction of the term $Z_{i}^T\pi^*$ to model invalid instruments and a fixed $\gamma^*$. 

Under the model in \eqref{eq:powerSetup}, we study whether a particular test has the power to detect the alternative $H_a: \beta^* \neq \beta_0; \pi_{B^C}^* \neq 0$ under the null $H_0: \beta^* = \beta_0; \pi_{B^C}^* = 0$ for a given set $B$. The first alternative is the power to detect $\beta^* \neq \beta_0$. The second alternative, perhaps a more important alternative in the case of invalid instruments, is power to detect the wrong $B$s in the union of $C_{1-\alpha}(Y,D,Z)$. A wrong $B$ is where $B$ does not contain all the invalid instruments so that $\pi_{B^C}^* \neq 0$. If a test has good power against the second alternative, we would be less likely to take the unions over wrong $B$s in $C_{1-\alpha}(Y,D,Z)$ and our robust confidence interval will tend to be shorter. We refer to analyzing power of tests under these two alternatives as power under invalid instruments. 

Under this invalid power framework, we can study the power of common test statistics in the instrumental variables literature. As an illustrative example, we study the power of the Anderson-Rubin test \citep{anderson_estimation_1949}; see the supplementary materials for additional analysis of different test statistics. The Anderson-Rubin test is a popular test in instrumental variables based on the partial F-test of the regression coefficients between $Y - D\beta_0$ versus $Z_{B^C}$, i.e.
\begin{equation} \label{eq:AR}
\text{\AR}(\beta_0,B) = \frac{ (Y - D \beta_0)^T (P_Z - P_{Z_B})(Y  - D \beta_0) / L - c(B)}{ (Y - D \beta_0)^T R_{Z} (Y - D \beta_0) / (n - L)}
\end{equation}
The Anderson-Rubin test has some attractive properties, including robustness to weak instruments (i.e. violation of (A1)) \citep{staiger_instrumental_1997}, simple formula, robustness to first-stage modeling assumptions, exact null distribution under Normality, and various others; see \citet{dufour_identification_2003} for details. A caveat to the Anderson-Rubin is its conservative power compared to a few recent tests \citep{andrews_optimal_2006, mikusheva_robust_2010}, such as the conditional likelihood ratio test \citep{moreira_conditional_2003}, under weak instruments. However, the Anderson-Rubin test has the feature that it can be used as a pretest to check whether the candidate subset of instruments $B$ contains all the invalid instruments \citep{kleibergen_generalizing_2007}. This feature is particularly useful for our problem where we have possibly invalid instruments. Indeed, as we will show in Theorem \ref{thm:powerAR} and in Section \ref{sec:simGeneral}, contrary to the weak instrument literature, the Anderson-Rubin test actually performs better than the conditional likelihood ratio test under invalid instruments.

\begin{theorem} \label{thm:powerAR} Consider any set $B \subset \{1,\ldots,L\}$ with $c(B) = U - 1$ and the null hypothesis $H_0: \beta^* = \beta_0$ and $B^C$ contains valid instruments versus the alternative $H_a: \beta^* \neq \beta_0$ or $B^C$ contains some invalid instruments. Under the data generating model in \eqref{eq:powerSetup}, the exact power of $\text{AR}(\beta_0,B)$ under invalid instruments is 
\begin{equation} \label{eq:powerAR}
\text{pr}\{\AR(\beta_0,B) \geq q_{1-\alpha}^{F_{L - c(B),n-L,0}} \} = 1 - F_{L - c(B), n - L,\eta(B)}(q_{1-\alpha}^{F_{L - c(B),n-L,0}})
\end{equation}
where $q_{1-\alpha}^{F_{L - c(B),n-L,\eta(B)}}$ is the $1 -\alpha$ quantile of the non-central F distribution with degrees of freedom $L - c(B)$, $n - L$ and non-centrality $\eta(B) = ||R_{Z_B} Z_{B^C}(\pi_{B^C}^* + \gamma_{B^C}^* (\beta^* - \beta_0)) ||_2^2$ 
\end{theorem}
The power of the Anderson-Rubin test under invalid instruments is a generalization of the power of the Anderson-Rubin when all the instruments are valid. Specifically, if $B$ contains all the invalid instruments so that $B^* \subseteq B $, then $\pi_{B^C}^* = 0$ and the non-centrality parameter $\eta(B)$ in Theorem \ref{thm:powerAR} would consist of only the strength of the instruments, specifically $||R_{Z_B} Z_{B^C} \gamma_{B^C} (\beta^* - \beta_0)) ||_2^2$, and we would return to the usual power of the Anderson-Rubin test with all valid instruments. On the other hand, if $B$ does not contain all invalid instruments so that $B^* \not\subseteq B$, then $\pi_{B^C}^* \neq 0$, the Anderson-Rubin test will still have power, even if $\beta^* = \beta_0$. That is, the Anderson-Rubin test will reject $H_0$ and will generally have shorter intervals when $B$ does not contain all the invalid instruments; see Section \ref{sec:simGeneral} for empirical verification of this phenomena. Also, the Anderson-Rubin has no power when $\pi_{B^C}* +  \gamma_{B^C}^* (\beta^* - \beta_0) = 0$; a similar result was shown in \citet{kadane_comment_1977} and \citet{small_sensitivity_2007} when studying the power of overidentifying restrictions tests. Finally, we note that our power formula is exact and does not invoke asymptotics. 

While there are many other ways to study the power of a test under invalid instruments, we believe the framework, based partly on the weak instrument literature, provides a first-step approximation to the behaviors of common tests under invalid instruments. The supplementary materials elaborates on this by (i) showing that the framework is a decent approximation to the invalid instrument phenomena, (ii) providing an asymptotic version of this framework under invalid instruments, (iii) analysis of other test statistics under this framework, including the two-stage least squares test, and (iv) studying the resulting power of the confidence interval $C_{1-\alpha}(Y,D,Z)$ where we take unions over all sets $B$.

\section{Simulation Study With Invalid Instruments} \label{sec:simGeneral}
\subsection{Coverage} \label{sec:simRobust}
In this simulation study, we evaluate the robustness of our method compared to popular methods for confidence intervals in the instrumental variables when there are invalid instruments. The simulation setup follows \eqref{eq:powerSetup} with $n = 5000$ individuals with $L = 10$ candidate instruments where each pair of instruments are correlated with correlation $0.6$.  We fix the parameters $\beta^* = 2$, $\sigma_2 = \sigma_1 = 1$, $\rho = 0.8$, and $U = 5$ (i.e. at most 50\% of the instruments are valid). We vary the parameters $\pi^*$ and $\gamma^*$ as follows. We change $\pi^*$'s support changes from $0$ to $4$ to represent invalid instruments and the elements of $\pi^*$ in the support are drawn from a uniform distribution. For $\gamma^*$, we set it two values that correspond to the concentration parameters $100$ and $5$ to represent strong and weak instruments. The concentration parameter is the expected value of the F statistic for the coefficients $Z_{(B^*)^C}$ in the regression of $D$ and $Z$ and is a measure of instrument strength \citep{stock_survey_2002}. For each simulation setting, we generate $5000$ simulated data sets.

For each simulation setting, we compare our methods in \eqref{eq:confIntProposal} and \eqref{eq:confIntProposalScreen} to ``naive'' and ``oracle'' methods where ``naive'' methods assume all candidate instruments are valid, which is typically done in practice, and ``oracle'' methods assume one knows exactly which instruments are valid and invalid, i.e. $B^*$ is known, and use \eqref{eq:confIntInversion}. Note that the oracle methods are not practical because of the incomplete knowledge about exactly which instruments are invalid versus valid. We use the following test statistics popular in instrumental variables literature, the two-stage least squares test, the Anderson-Rubin test in \eqref{eq:AR}, and the conditional likelihood ratio test \citep{moreira_conditional_2003}. Also, for our methods involving pretests in \eqref{eq:confIntProposalScreen}, we use the Sargan test as the pretests for the two-stage least squares test and the conditional likelihood ratio test, both at level $\alpha_1 = 0.01$ for the pretest, and $\alpha_2 = 0.04$ for the subsequent tests. For simulation settings involving weak instruments, we only compare between tests that are robust to weak instruments, specifically the the Anderson-Rubin test and the conditional likelihood ratio test, and we do not use the pretesting method in \eqref{eq:confIntProposalScreen}, which uses the Sargan test and is known to perform poorly with weak instruments \citep{staiger_instrumental_1997}; see the Supplementary materials for details about the tests and additional details about the simulation setting.

Table \ref{tab:1} shows the empirical coverage proportion of different methods when we vary $s$ and the strength of the instruments. When there are no invalid instruments, $s = 0$, the naive and oracle procedures have the desired 95\% coverage for both strong and weak instruments. Our methods have higher than 95\% coverage because they do not assume that all candidate instruments are valid. As the number of invalid instruments, $s$, increases, the naive methods fail to have any coverage regardless of the strength of the instruments. Our methods, in contrast, have the desired level of coverage, with the coverage level reaching nominal levels when $s$ is at the boundary of $s < U$, i.e. $s = 4$, all without knowing which instruments are valid or invalid a priori. The oracle methods have coverage reaching nominal levels since they know which instruments are valid and invalid. Finally, we note that under the worst case, where the instruments are weak and $0 < s < U$ so that essentially, all three instrumental variables assumptions (A1)-(A3) are violated, our methods still provide honest coverage and when $s = 4$, reach oracle coverage.

In short, in the presence of possibly invalid instruments, the naive, popular approach of simply assuming all the instruments are valid would lead to misleading inference. In contrast, our methods provide honest coverage regardless of whether instruments are invalid or valid and should be used whenever there is concern for possibly invalid instruments. In particular, \eqref{eq:confIntProposal} works regardless of the strength of the instruments while our method in \eqref{eq:confIntProposalScreen} provides a desired level of coverage so long as the instruments are strong.
\begin{table}
\def~{\hphantom{0}}
\tbl{Comparison of coverage between 95\% confidence intervals}{
\begin{tabular}{l l lccccc}
 \\
Strength &                                            Case &         Test & $s = 0$ & $s = 1$ & $s = 2$ & $s = 3$ & $s = 4$ \\[5pt]
Strong   &                                           Naive &         TSLS &     ~94 &     ~~0 &     ~~0 &     ~~0 &    ~~0 \\
         &                                                 &           AR &     ~95 &     ~~0 &     ~~0 &     ~~0 &    ~~0 \\
         &                                                 &          CLR &     ~95 &     ~~0 &     ~~0 &     ~~0 &    ~~0 \\
         &                                      Our method &         TSLS &     100 &     100 &     100 &     100 &    ~94 \\
         &                                                 &           AR &     100 &     100 &     100 &     100 &    ~95 \\
         &                                                 &          CLR &     100 &     100 &     100 &     100 &    ~98 \\
         &                                                 & SAR $+$ TSLS &     100 &     100 &     100 &     100 &    ~95 \\
         &                                                 &  SAR $+$ CLR &     100 &     100 &     100 &     100 &    ~96 \\ 
         &                                          Oracle &         TSLS &     ~94 &     ~94 &     ~94 &     ~94 &    ~94 \\
         &                                                 &           AR &     ~95 &     ~95 &     ~95 &     ~95 &    ~95 \\
         &                                                 &          CLR &     ~95 &     ~95 &     ~95 &     ~95 &    ~96 \\
Weak &                                  Naive    &           AR &     ~95 &     ~~0 &     ~~0 &     ~~0 &    ~~0 \\
         &                                                 &          CLR &     ~95 &     ~~0 &     ~~0 &     ~~0 &    ~~0 \\
         &                            Our method  &           AR &     100 &     100 &     100 &     100 &    ~95 \\
         &                                                 &          CLR &     100 &     100 &     100 &     100 &    ~95 \\
         &                            Oracle          &           AR &     ~95 &     ~95 &     ~95 &     ~95 &    ~95 \\
         &                                                 &          CLR &     ~95 &     ~95 &     ~95 &     ~95 &    ~95 
\end{tabular}}
\label{tab:1}
\begin{tabnote}
TSLS, two-stage least squares; AR, Anderson--Rubin test; CLR, conditional likelihood ratio test; SAR, Sargan test. There are $L = 10$ candidate instruments and $U = 5$. Strong instruments and weak instruments correspond to concentration parameters equaling 100 and 5, respectively. The standard error for all the coverage proportions do not exceed 1\%.
\end{tabnote}
\end{table}
 
\subsection{Length and power} \label{sec:simInformative}
Next, following Section \ref{sec:power}, we compare tests that are used in our methods to see which tests are ``better'' with regards to the length and power of the corresponding confidence interval. The simulation settings are identical to Section \ref{sec:simRobust}, except we only compare between our methods and the oracles. Table \ref{tab:2} examines the median length of the 95\% confidence intervals.

\begin{table}
\def~{\hphantom{0}}
\tbl{Comparison of median lengths between 95\% confidence intervals}{
\begin{tabular}{l l lccccc}
 \\
Strength &           Case  &                  Test &            $s = 0$ &            $s = 1$ &         $s = 2$ &         $s = 3$ &   ~$s = 4$ \\[5pt]
Strong    & Our method &                TSLS &    $0 \cdot 93$ &  ~$2 \cdot 63$ & $2 \cdot 08$ & $3 \cdot 62$ & ~~$5 \cdot 12$ \\
              &                      &                    AR &    $1 \cdot 63$ &  ~$0 \cdot 77$ & $0 \cdot 51$ & $0 \cdot 36$ &  ~~$0 \cdot 24$ \\
              &                      &                  CLR &    $1 \cdot 09$ &  $12 \cdot 42$ & $8 \cdot 89$ & $9 \cdot 44$ &  ~~$8 \cdot 02$ \\
              &                      & SAR $+$ TSLS &    $0 \cdot 95$ &  ~$0 \cdot 54$ & $0 \cdot 37$ & $0 \cdot 26$ & ~~$0 \cdot 17$ \\
              &                      &  SAR $+$ CLR &    $1 \cdot 12$ &  ~$0 \cdot 58$ & $0 \cdot 38$ & $0 \cdot 27$ &  ~~$0 \cdot 17$ \\ 
              &           Oracle &               TSLS &    $0 \cdot 12$ &  ~$0 \cdot 13$ & $0 \cdot 14$ & $0 \cdot 15$ &  ~~$0 \cdot 16$ \\
              &                      &                    AR &    $0 \cdot 20$ &  ~$0 \cdot 21$ & $0 \cdot 21$ & $0 \cdot 22$ &  ~~$0 \cdot 24$ \\
              &                      &                  CLR &    $0 \cdot 12$ &  ~$0 \cdot 13$ & $0 \cdot 14$ & $0 \cdot 15$ &  ~~$0 \cdot 16$ \\
Weak     &  Our method &                    AR &      ~~~$\infty$ &      ~~~$\infty$ &      ~~$\infty$ &    ~~$\infty$ & $573 \cdot 01$   \\
              &                      &                  CLR &      ~~~$\infty$ &      ~~~$\infty$ &      ~~$\infty$ &    ~~$\infty$ &  ~~~$\infty$   \\
              &           Oracle &                    AR &    $1 \cdot 02$ & ~$ 1 \cdot 07$ &   $1 \cdot13$ &$1 \cdot 19$ &  ~~$1 \cdot 27$ \\
              &                       &                 CLR &    $0 \cdot 62$ &  ~$0 \cdot 65$ &  $0 \cdot 70$ & $0 \cdot 75$ &  ~~$0 \cdot 82$ 
\end{tabular}}
\label{tab:2}
\begin{tabnote}
TSLS, two-stage least squares; AR, Anderson--Rubin test; CLR, conditional likelihood ratio test; SAR, Sargan test. There are $L = 10$ candidate instruments and $U = 5$. Strong instruments and weak instruments correspond to concentration parameters equaling 100 and 5, respectively. The interquartile range of all strong intervals do not exceed $5 \cdot 77$. The interquartile range of weak intervals do not exceed $1 \cdot 00$ for non-infinite intervals except when $s = 4$.
\end{tabnote}
\end{table}

For strong instruments, our method and the oracles become similar as the number of invalid instruments $s$ grows, with the Anderson-Rubin test and methods with pretesting achieving oracle performance $s = 4$ while the two-stage least squares and conditional likelihood ratio tests, both without pretesting, not reaching oracle performance at $s = 4$. The improved performance with pretesting is expected since pretesting was introduced to remove taking unnecessary unions of intervals in \eqref{eq:confIntProposalScreen}. For weak instruments, our method produces infinite intervals with the exception of the Anderson-Rubin test at $s = 4$. The infinite lengths suggest that weak instruments can greatly amplify the bias caused by invalid instruments, thereby forcing our robust methods to produce infinite intervals to retain honest coverage, something that has been observed in previous studies \citep{small_war_2008}. In contrast, the oracle intervals produce finite intervals since instrumental validity is not an issue; although if the instrument is arbitrary weak, infinite confidence intervals are necessary \citep{dufour_impossibility_1997}.

\begin{figure}
\figurebox{20pc}{32pc}{}[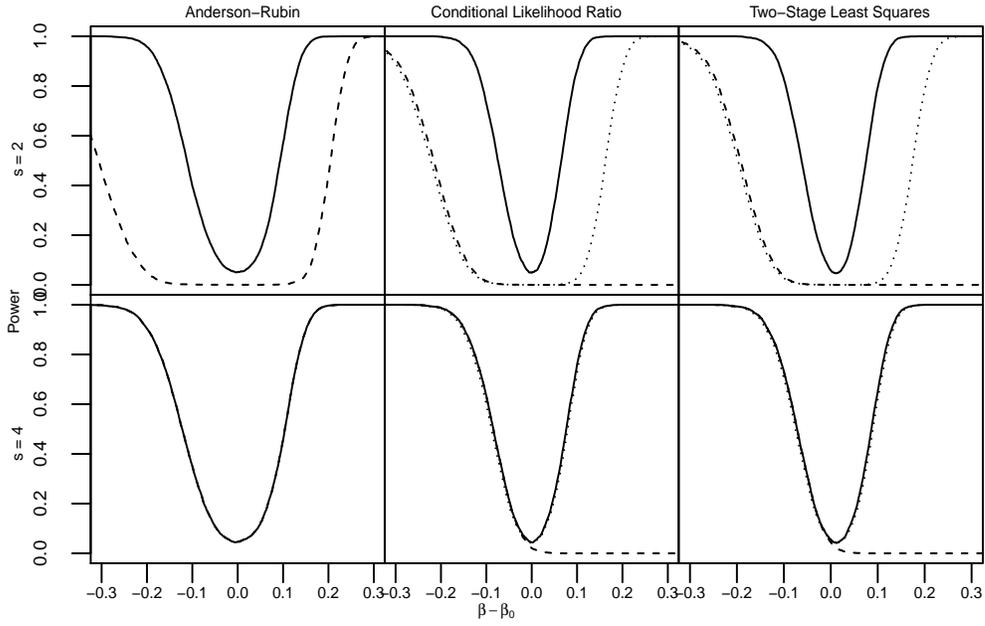]
\caption{Power curves between our methods and the oracle methods for strong instrument. Oracles intervals are solid lines. Dashed lines are our methods. Dotted lines are pretesting methods. Each column represents each test statistic while each row represents different $s$.}
\label{fig:powerStrongIV}
\end{figure}

\begin{figure}
\figurebox{20pc}{32pc}{}[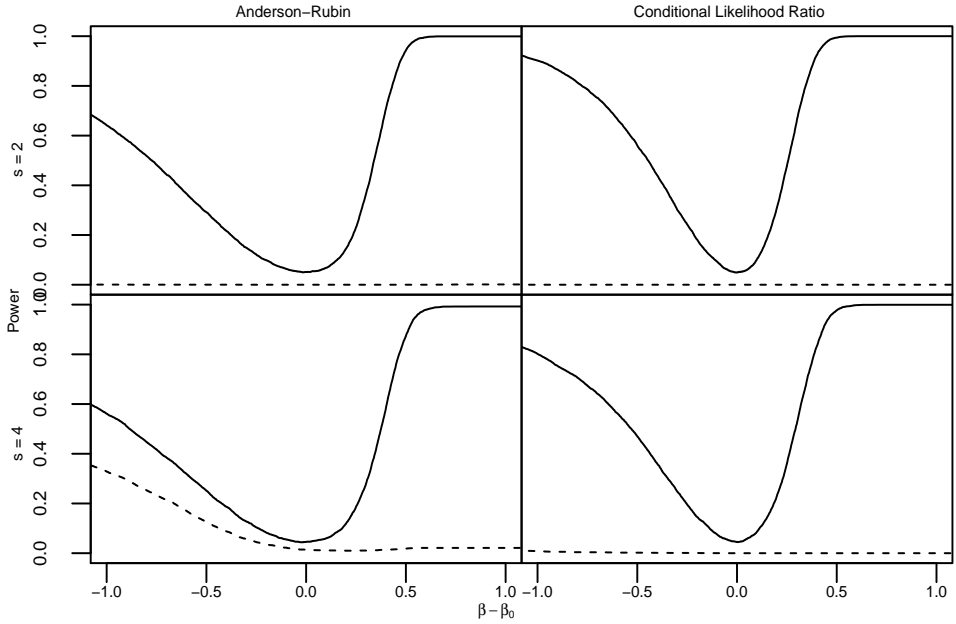]
\caption{Power curves between our methods and the oracle methods for weak instrument. Circles represent the Anderson-Rubin tests, triangles represent the conditional likelihood ratio test, and squares represent the two-stage least squares. Solid lines represent oracle methods. Dashed lines represent our methods. Dotted lines represent pretesting methods.}
\label{fig:powerWeakIV}
\end{figure}

Figures \ref{fig:powerStrongIV} and \ref{fig:powerWeakIV} compares the power of the tests under strong and weak instruments, respectively. Each column represents each tests while each row represents different $s$.

For strong instruments in Figure \ref{fig:powerStrongIV}, all our methods' power curves are dominated by the oracles' power curves, which is to be expected since the oracles know exactly which instruments are valid and invalid. Similar to what we observed with confidence interval length, the gap between our methods and the oracle shrinks as $s$ grows. Also, two-stage least squares and the conditional likelihood ratio test without pretesting have no power for the positive side of the alternative close to zero while the pretesting versions provide power in this region. The Anderson-Rubin test, which does not require a pretest, has power on both sides of the alternative and at $s = 4$, the Anderson-Rubin achieves oracle power with perfect overlapping curves.

For weak instruments in Figure \ref{fig:powerWeakIV}, again, our methods' power curves are dominated by the oracles' power curves. Between the two tests for our methods, the Anderson-Rubin test performs no worse than the conditional likelihood ratio test, with the Anderson-Rubin test dominating the conditional likelihood ratio test for $s = 4$. This finding contrasts with the advice in weak instruments literature where the conditional likelihood ratio generally dominates the Anderson-Rubin test \citep{andrews_optimal_2006, mikusheva_robust_2010}; the weak instrument setting suggests that the intuition from the weak instrument literature with regards to power does not necessary translate if invalid instruments are present.

In short, the simulation study here suggests (i) there may not be uniformly dominating method or tests used with our methods (ii) the performance of the methods depend on $s$ and the strength of the instruments, and that (iii) invalid instruments may exacerbate the bias from weak instruments and traditionally strong performers in weak instrument settings, like conditional likelihood ratio test, may perform poorly compared to the Anderson-Rubin test, traditionally a weak performer in weak instruments. For practice, when one has strong instruments, the Anderson-Rubin test and the pretesting method with two-stage least squares or conditional likelihood ratio test perform well with respect to power and length, with Anderson-Rubin test being the simpler one due to its lack of a pretest but the pretesting method providing somewhat shorter length. For weak instruments, the Anderson-Rubin test performs the best. Indeed, in this worst case where the instruments are weak and there are invalid instruments (i.e. all instrumental variables assumptions are violated), all our procedures leads to infinite, but honest, intervals. The presence of an infinite interval can be disappointing at first, but it's actually informative in the sense that the data's quality is insufficient to draw any meaningful conclusion about the treatment effect $\beta^*$.

\section{Data Analysis: Demand for Food in Developmental Economics}

In developmental economics, there is great interest in studying relationships between income and expenditures on goods and services to better understand welfare policies and economic activity in developing nations \citep{deaton_analysis_1997}. There is a long-held hypothesis, loosely called the efficient wage hypothesis, which suggests that raising income would lead to workers being better fed, which in turn lead to better productivity in the workforce, especially in developing economies \citep{mirrlees_pure_1975,stiglitz_efficiency_1976, subramanian_demand_1996}. Consequently, a strong focus in this literature is looking at the relationship between income and calorie intake \citep{reutlinger_malnutrition_1976,behrman_will_1987,ravallion_income_1990,bouis_agricultural_1990,bouis_estimates_1992,subramanian_demand_1996,dawson_estimating_1998,tiffin_demand_2002,abdulai_nonparametric_2004}.

To this end, we reanalyze the instrumental variables analysis done in \citet{bouis_agricultural_1990} and \citet{bouis_estimates_1992} where the goal was to analyze the effect of income on demand for food among $n = 405$ Philippine farm households. Specifically, the outcome is household food expenditures, $Y_i$. The exposure is the household's log income, $D_i$. We use four candidate instruments, cultivated area per capita, $Z_{i1}$, worth of assets, $Z_{i2}$, binary indicator of electricity at the household, $Z_{i3}$, and quality of flooring at the house, $Z_{i4}$. Page 82 of \citet{bouis_agricultural_1990} states that the reasoning behind proposing these variables as instrumental variables is that ``land availability is assumed to be a constraint in the short run, and therefore exogenous to the household decision making process.'' We also control for the measured covariates, which are mother's education, father's education, mother's age, father's age, mother's nutritional knowledge, price of corn, price of rice, population density of the municipality, and number of household members in adult equivalents; see \citet{bouis_agricultural_1990} and \citet{bouis_estimates_1992} for further details on the data.

The F-statistic for instrument strength is 103$\cdot$77, indicating reasonably strong instruments. The Sargan test for overidentification, which tests assumptions (A2) and (A3), produces a p-value of 0$\cdot$079. Even though the p-value is low, usually practitioners of the instrumental variables method would assume (A2) and (A3) are true since the p-value is above 0$\cdot$0.5, the typical threshold for significance level and use one of the four naive methods in Section \ref{sec:simGeneral} to obtain confidence intervals; see column $U = 1$ in Table \ref{tab:data}, row with Sargan pre-testing as examples of this type of analysis. In contrast, our methods do not take for granted that the four instruments are valid and allow the instruments to be invalid. In particular, we vary $U$ away from $1$ until the confidence intervals from our methods contain the null value $0$ and use the three tests in Section \ref{sec:simRobust}. For procedures with pretests, we used the same $\alpha_1$ and $\alpha_2$ we did in Section \ref{sec:simRobust}. 

The empirical findings are summarized in Table \ref{tab:data}. Even if there is an instrument that is invalid, there is a significant effect of income on food expenditures; at $U = 2$, all the tests do not contain the null value zero. But, if more than one instrument is invalid, $U > 2$, all the tests contain the null value zero and the causal effect is no longer significant. We also note that our method in \eqref{eq:confIntProposal} using the Anderson--Rubin or the pre-testing method with two-stage least squares provide the shortest interval.

\begin{table}
\tbl{95\% Confidence Interval of Income's Effect on Food Expenditures}{
\begin{tabular}{l c c c}
 \\
 Test &  $U = 1$ (Naive) & $U = 2$ & $U = 3$ \\[5pt]
TSLS & (  0 $\cdot$ 043,~0 $\cdot$ 053) &  (  0 $\cdot$ 031,~0 $\cdot$ 059) & (-0$\cdot$017,~0$\cdot$064)\\
AR & (  0 $\cdot$ 044,~0 $\cdot$ 054) &  (  0 $\cdot$ 037,~0 $\cdot$ 058) & (-0$\cdot$027,~0$\cdot$068) \\
 CLR & (  0 $\cdot$ 043,~0 $\cdot$ 055) & (  0 $\cdot$ 034,~0 $\cdot$ 066) & (-0$\cdot$042,~0$\cdot$070)\\
SAR $+$ TSLS &  (0 $\cdot$ 042,~0 $\cdot$ 054) & (  0 $\cdot$ 031,~0 $\cdot$ 059) & (-0$\cdot$018,~0$\cdot$065) \\
 SAR $+$ CLR & (0 $\cdot$ 043,~0 $\cdot$ 055) &(  0 $\cdot$ 034,~0 $\cdot$ 067) & (-0$\cdot$049,~0$\cdot$072) 
\end{tabular}}
\label{tab:data}
\begin{tabnote}
TSLS, two-stage least squares; AR, Anderson--Rubin test; CLR, conditional likelihood ratio test; SAR, Sargan test. There are four candidate instruments.
\end{tabnote}
\end{table}
%
%
%
%

%
%
%
%

The empirical application illustrates the usefulness of our procedure whenever there is a concern for invalid instruments. In particular, our procedure is a simple modification of pre-existing procedures for instrumental variables which yield honest confidence intervals and can act as a sensitivity analysis for violation of IV assumptions (A2) and (A3). The supplementary materials provides additional data analysis with this data to further highlight the strengths of our confidence interval approach in practice.

\section{Discussion}

This paper proposes a simple and general method to construct robust confidence intervals for causal effects using instrumental variables estimates when the instruments are possibly invalid, with theoretical guarantees with respect to coverage. We propose two methods in \eqref{eq:confIntProposal} and \eqref{eq:confIntProposalScreen} that are simple modifications of pre-existing methods in instrumental variables that protects against invalid instruments. Our data analysis example illustrates that our method can be a simple, robust alternative to confidence intervals that have the proper coverage whenever there is concern for possibly invalid instruments and can assess the sensitivity of our inference to violations of IV assumptions. 

\section*{Supplementary material}
\label{SM}
Supplementary material available at \Bka\ online includes theoretical details, additional simulations and empirical analysis.

\appendix

\appendixone
\section*{Appendix 1}
\subsection{Proofs}
\begin{proof}[of Theorem~\ref{theorem1}]
By $s = c(B^*) < U$, there is a subset $\tilde{B}$ where $c(\tilde{B}) = U-1$ and $B^* \subseteq \tilde{B}$. Also, its complement $\tilde{B}^C$ only contains valid instruments and thus, $\text{pr}\{\beta^* \in C_{1 - \alpha}(Y,D,Z,\tilde{B}) \} \geq 1 - \alpha$. Hence, we have 
\[
\text{pr}\{\beta^* \in C_{1 - \alpha}(Y,D,Z)\} \geq \text{pr}\{\beta^* \in C_{1- \alpha}(Y,D,Z,\tilde{B})\} \geq 1- \alpha
\]
for all values of $\beta^*$.
\end{proof}

\begin{proof}[of Theorem~\ref{theorem2}] Similar to the proof for Theorem \ref{theorem1}, $\tilde{B}$, which is a superset containing all invalid instruments, has to exist and additionally, have the property $\text{pr}\{S(\tilde{B}) \geq q_{1 - \alpha_1}\} \leq \alpha_1$. Then, we can use Bonferroni's inequality to obtain
\begin{align*}
\text{pr}\{\beta^* \in C_{1 - \alpha}'(Y,D,Z)\} &\geq \text{pr}\{\beta^* \in C_{1- \alpha_2}(Y,D,Z,\tilde{B}) \cap S(\tilde{B}) \leq q_{1 - \alpha_1} \} \\
&\geq 1 - \text{pr}\{\beta^* \notin C_{1- \alpha_2}(Y,D,Z,\tilde{B})\} - \text{pr}\{S(\tilde{B}) \geq q_{1 - \alpha_1} \} \\
&\geq 1 - \alpha_1 - \alpha_2 = 1 - \alpha
\end{align*}
thereby guaranteeing the correct coverage.
\end{proof}

\begin{proof}[of Theorem~\ref{thm:powerAR}] 
By Cochran's theorem, (i) the numerator and the denominator of \eqref{eq:AR} are independent, (ii) the denominator, scaled by $\tilde{\sigma}^2 = \sigma_2^2 + (\beta^* - \beta_0)^2 \sigma_1^2 + 2(\beta^* - \beta_0)\rho \sigma_1 \sigma_2$, is a central chi-square with $n-L$ degrees of freedom, i.e.
\[
\frac{(Y - D \beta_0)^T R_{Z} (Y - D \beta_0)}{\tilde{\sigma}^2 (n - L)} = \frac{\{(\beta^* - \beta_0) \xi + \epsilon\}^T R_Z \{(\beta^* - \beta_0) \xi + \epsilon\}}{\tilde{\sigma}^2 (n - L)} \sim \chi_{n - L,0}^2
\]
and (iii), the numerator, scaled by $\tilde{\sigma}^2$, is a non-central chi-square distribution with non-centrality $\eta(B)$
\[
 \frac{ (Y - D \beta_0)^T (P_Z - P_{Z_B})(Y  - D \beta_0)}{\tilde{\sigma}^2 \{L - c(B)\}} \sim \chi_{L - c(B),\eta(B)}^2,\quad{} \eta(B) = \| (P_Z - P_{Z_B}) Z\{\pi^* + \gamma^*(\beta^* - \beta_0)\} \|_2^2
\]
Since $(P_Z - P_{Z_B})Z$ can be rewritten as the residual projection of $Z$ onto $Z_{B}$, i.e.
\[
(P_Z  - P_{Z_B})Z = Z - [Z_B : P_{Z_B} Z_{B^C}] = [0 : Z_{B^C} - P_{Z_B} Z_{B^C}] =[0 : R_{Z_B} Z_{B^C}] = R_{Z_B} Z
\]
the $\text{\AR}(\beta_0,B)$ is a non-central F distribution with degrees of freedom. 
\end{proof}

%
%
%
%
%

\bibliographystyle{biometrika}
\bibliography{mainbib}

\begin{thebibliography}{49}
\expandafter\ifx\csname natexlab\endcsname\relax\def\natexlab#1{#1}\fi

\bibitem[{Abdulai \& Aubert(2004)}]{abdulai_nonparametric_2004}
\textsc{Abdulai, A.} \& \textsc{Aubert, D.} (2004).
\newblock Nonparametric and parametric analysis of calorie consumption in
  tanzania.
\newblock \textit{Food Policy} \textbf{29}, 113 -- 129.

\bibitem[{Anderson \& Rubin(1949)}]{anderson_estimation_1949}
\textsc{Anderson, T.~W.} \& \textsc{Rubin, H.} (1949).
\newblock Estimation of the parameters of a single equation in a complete
  system of stochastic equations.
\newblock \textit{Annals of Mathematical Statistics} \textbf{20}, 46--63.

\bibitem[{Andrews \& Lu(2001)}]{andrews_consistent_2001}
\textsc{Andrews, D.~W.} \& \textsc{Lu, B.} (2001).
\newblock Consistent model and moment selection procedures for gmm estimation
  with application to dynamic panel data models.
\newblock \textit{Journal of Econometrics} \textbf{101}, 123--164.

\bibitem[{Andrews(1999)}]{andrews_consistent_1999}
\textsc{Andrews, D. W.~K.} (1999).
\newblock Consistent moment selection procedures for generalized method of
  moments estimation.
\newblock \textit{Econometrica} \textbf{67}, 543--563.

\bibitem[{Andrews et~al.(2006)Andrews, Moreira \& Stock}]{andrews_optimal_2006}
\textsc{Andrews, D. W.~K.}, \textsc{Moreira, M.~J.} \& \textsc{Stock, J.~H.}
  (2006).
\newblock Optimal two-sided invariant similar tests for instrumental variables
  regression.
\newblock \textit{Econometrica} \textbf{74}, 715--752.

\bibitem[{Andrews et~al.(2007)Andrews, Moreira \&
  Stock}]{andrews_performance_2007}
\textsc{Andrews, D. W.~K.}, \textsc{Moreira, M.~J.} \& \textsc{Stock, J.~H.}
  (2007).
\newblock Performance of conditional wald tests in \{IV\} regression with weak
  instruments.
\newblock \textit{Journal of Econometrics} \textbf{139}, 116--132.

\bibitem[{Angrist et~al.(1996)Angrist, Imbens \&
  Rubin}]{angrist_identification_1996}
\textsc{Angrist, J.~D.}, \textsc{Imbens, G.~W.} \& \textsc{Rubin, D.~B.}
  (1996).
\newblock Identification of causal effects using instrumental variables.
\newblock \textit{Journal of the American statistical Association} \textbf{91},
  444--455.

\bibitem[{Baiocchi et~al.(2014)Baiocchi, Cheng \&
  Small}]{baiocchi_instrumental_2014}
\textsc{Baiocchi, M.}, \textsc{Cheng, J.} \& \textsc{Small, D.~S.} (2014).
\newblock Instrumental variable methods for causal inference.
\newblock \textit{Statistics in Medicine} \textbf{33}, 2297--2340.

\bibitem[{Behrman \& Deolalikar(1987)}]{behrman_will_1987}
\textsc{Behrman, J.~R.} \& \textsc{Deolalikar, A.~B.} (1987).
\newblock Will developing country nutrition improve with income? a case study
  for rural south india.
\newblock \textit{Journal of political Economy} \textbf{95}, 492--507.

\bibitem[{Bouis \& Haddad(1990)}]{bouis_agricultural_1990}
\textsc{Bouis, H.~E.} \& \textsc{Haddad, L.~J.} (1990).
\newblock \textit{Agricultural Commercialization, Nutrition, and the Rural
  Poor}.
\newblock Lynne Rienner Publishers.

\bibitem[{Bouis \& Haddad(1992)}]{bouis_estimates_1992}
\textsc{Bouis, H.~E.} \& \textsc{Haddad, L.~J.} (1992).
\newblock Are estimates of calorie-income elasticities too high?: A
  recalibration of the plausible range.
\newblock \textit{Journal of Development Economics} \textbf{39}, 333--364.

\bibitem[{Bowden et~al.(2015)Bowden, Davey~Smith \&
  Burgess}]{bowden_mendelian_2015}
\textsc{Bowden, J.}, \textsc{Davey~Smith, G.} \& \textsc{Burgess, S.} (2015).
\newblock Mendelian randomization with invalid instruments: effect estimation
  and bias detection through egger regression.
\newblock \textit{International Journal of Epidemiology} \textbf{44}, 512--525.

\bibitem[{Burgess et~al.(2012)Burgess, Butterworth, Malarstig \&
  Thompson}]{burgess_use_2012}
\textsc{Burgess, S.}, \textsc{Butterworth, A.}, \textsc{Malarstig, A.} \&
  \textsc{Thompson, S.~G.} (2012).
\newblock Use of mendelian randomisation to assess potential benefit of
  clinical intervention.
\newblock \textit{British Medical Journal} \textbf{345}.

\bibitem[{Cheng \& Liao(2015)}]{cheng_select_2015}
\textsc{Cheng, X.} \& \textsc{Liao, Z.} (2015).
\newblock Select the valid and relevant moments: An information-based lasso for
  gmm with many moments.
\newblock \textit{Journal of Econometrics} \textbf{186}, 443--464.

\bibitem[{Conley et~al.(2012)Conley, Hansen \& Rossi}]{conley_plausibly_2012}
\textsc{Conley, T.~G.}, \textsc{Hansen, C.~B.} \& \textsc{Rossi, P.~E.} (2012).
\newblock Plausibly exogenous.
\newblock \textit{Review of Economics and Statistics} \textbf{94}, 260--272.

\bibitem[{Davey~Smith \& Ebrahim(2003)}]{davey_smith_mendelian_2003}
\textsc{Davey~Smith, G.} \& \textsc{Ebrahim, S.} (2003).
\newblock €˜mendelian randomization': can genetic epidemiology contribute to
  understanding environmental determinants of disease?
\newblock \textit{International Journal of Epidemiology} \textbf{32}, 1--22.

\bibitem[{Davey~Smith \& Ebrahim(2004)}]{davey_smith_mendelian_2004}
\textsc{Davey~Smith, G.} \& \textsc{Ebrahim, S.} (2004).
\newblock Mendelian randomization: prospects, potentials, and limitations.
\newblock \textit{International Journal of Epidemiology} \textbf{33}, 30--42.

\bibitem[{Davidson \& MacKinnon(1993)}]{davidson_estimation_1993}
\textsc{Davidson, R.} \& \textsc{MacKinnon, J.~G.} (1993).
\newblock \textit{Estimation and Inference in Econometrics}.
\newblock New York: Oxford University Press.

\bibitem[{Dawson \& Tiffin(1998)}]{dawson_estimating_1998}
\textsc{Dawson, P.~J.} \& \textsc{Tiffin, R.} (1998).
\newblock Estimating the demand for calories in india.
\newblock \textit{American Journal of Agricultural Economics} \textbf{80},
  474--481.

\bibitem[{Deaton(1997)}]{deaton_analysis_1997}
\textsc{Deaton, A.} (1997).
\newblock \textit{The analysis of household surveys: a microeconometric
  approach to development policy}.
\newblock World Bank Publications.

\bibitem[{Dufour(1997)}]{dufour_impossibility_1997}
\textsc{Dufour, J.-M.} (1997).
\newblock Some impossibility theorems in econometrics with applications to
  structural and dynamic models.
\newblock \textit{Econometrica} , 1365--1387.

\bibitem[{Dufour(2003)}]{dufour_identification_2003}
\textsc{Dufour, J.-M.} (2003).
\newblock Identification, weak instruments, and statistical inference in
  econometrics.
\newblock \textit{The Canadian Journal of Economics / Revue canadienne
  d'Economique} \textbf{36}, 767--808.

\bibitem[{Hansen(1982)}]{hansen_large_1982}
\textsc{Hansen, L.~P.} (1982).
\newblock Large sample properties of generalized method of moments estimators.
\newblock \textit{Econometrica: Journal of the Econometric Society} ,
  1029--1054.

\bibitem[{Holland(1988)}]{holland_causal_1988}
\textsc{Holland, P.~W.} (1988).
\newblock Causal inference, path analysis, and recursive structural equations
  models.
\newblock \textit{Sociological Methodology} \textbf{18}, 449--484.

\bibitem[{Kadane \& Anderson(1977)}]{kadane_comment_1977}
\textsc{Kadane, J.~B.} \& \textsc{Anderson, T.~W.} (1977).
\newblock A comment on the test of overidentifying restrictions.
\newblock \textit{Econometrica} \textbf{45}, 1027--1031.

\bibitem[{Kang et~al.(2015)Kang, Zhang, Cai \& Small}]{kang_instrumental_2015}
\textsc{Kang, H.}, \textsc{Zhang, A.}, \textsc{Cai, T.~T.} \& \textsc{Small,
  D.~S.} (2015).
\newblock Instrumental variables estimation with some invalid instruments and
  its application to mendelian randomization.
\newblock \textit{Journal of the American Statistical Association} , to appear.

\bibitem[{Kleibergen(2007)}]{kleibergen_generalizing_2007}
\textsc{Kleibergen, F.} (2007).
\newblock Generalizing weak instrument robust iv statistics towards multiple
  parameters, unrestricted covariance matrices and identification statistics.
\newblock \textit{Journal of Econometrics} \textbf{139}, 181--216.

\bibitem[{Koles{\'a}r et~al.(2013)Koles{\'a}r, Chetty, Friedman, Glaeser \&
  Imbens}]{kolesar_identification_2013}
\textsc{Koles{\'a}r, M.}, \textsc{Chetty, R.}, \textsc{Friedman, J.~N.},
  \textsc{Glaeser, E.~L.} \& \textsc{Imbens, G.~W.} (2013).
\newblock Identification and inference with many invalid instruments.
\newblock \textit{National Bureau of Economic Research} , No. w17519.

\bibitem[{Lawlor et~al.(2008)Lawlor, Harbord, Sterne, Timpson \&
  Davey~Smith}]{lawlor_mendelian_2008}
\textsc{Lawlor, D.~A.}, \textsc{Harbord, R.~M.}, \textsc{Sterne, J. A.~C.},
  \textsc{Timpson, N.} \& \textsc{Davey~Smith, G.} (2008).
\newblock Mendelian randomization: Using genes as instruments for making causal
  inferences in epidemiology.
\newblock \textit{Statistics in Medicine} \textbf{27}, 1133--1163.

\bibitem[{Liao(2013)}]{liao_adaptive_2013}
\textsc{Liao, Z.} (2013).
\newblock Adaptive gmm shrinkage estimation with consistent moment selection.
\newblock \textit{Econometric Theory} \textbf{29}, 857--904.

\bibitem[{Mikusheva(2010)}]{mikusheva_robust_2010}
\textsc{Mikusheva, A.} (2010).
\newblock Robust confidence sets in the presence of weak instruments.
\newblock \textit{Journal of Econometrics} \textbf{157}, 236--247.

\bibitem[{Mirrlees(1975)}]{mirrlees_pure_1975}
\textsc{Mirrlees, J.} (1975).
\newblock A pure theory of underdeveloped economies.
\newblock \textit{Agriculture in development theory} \textbf{4}, 84--108.

\bibitem[{Moreira(2003)}]{moreira_conditional_2003}
\textsc{Moreira, M.~J.} (2003).
\newblock A conditional likelihood ratio test for structural models.
\newblock \textit{Econometrica} \textbf{71}, 1027--1048.

\bibitem[{Murray(2006)}]{murray_avoiding_2006}
\textsc{Murray, M.~P.} (2006).
\newblock Avoiding invalid instruments and coping with weak instruments.
\newblock \textit{The Journal of Economic Perspectives} \textbf{20}, 111--132.

\bibitem[{Ravallion(1990)}]{ravallion_income_1990}
\textsc{Ravallion, M.} (1990).
\newblock Income effects on undernutrition.
\newblock \textit{Economic development and cultural change} \textbf{38},
  489--515.

\bibitem[{Reutlinger \& Selowsky(1976)}]{reutlinger_malnutrition_1976}
\textsc{Reutlinger, S.} \& \textsc{Selowsky, M.} (1976).
\newblock \textit{Malnutrition and poverty. Magnitude and policy options.}
\newblock Baltimore: World Bank, Johns Hopkins University Press.

\bibitem[{Rosenbaum(2002)}]{rosenbaum_observational_2002}
\textsc{Rosenbaum, P.~R.} (2002).
\newblock \textit{Observational Studies}.
\newblock Springer Series in Statistics. Springer-Verlag, New York, 2nd ed.

\bibitem[{Rubin(1974)}]{rubin_estimating_1974}
\textsc{Rubin, D.~B.} (1974).
\newblock Estimating causal effects of treatments in randomized and
  nonrandomized studies.
\newblock \textit{Journal of Educational Psychology} \textbf{66}, 688.

\bibitem[{Sargan(1958)}]{sargan_estimation_1958}
\textsc{Sargan, J.~D.} (1958).
\newblock The estimation of economic relationships using instrumental
  variables.
\newblock \textit{Econometrica} , 393--415.

\bibitem[{Small(2007)}]{small_sensitivity_2007}
\textsc{Small, D.~S.} (2007).
\newblock Sensitivity analysis for instrumental variables regression with
  overidentifying restrictions.
\newblock \textit{Journal of the American Statistical Association}
  \textbf{102}, 1049--1058.

\bibitem[{Small \& Rosenbaum(2008)}]{small_war_2008}
\textsc{Small, D.~S.} \& \textsc{Rosenbaum, P.~R.} (2008).
\newblock War and wages: the strength of instrumental variables and their
  sensitivity to unobserved biases.
\newblock \textit{Journal of the American Statistical Association}
  \textbf{103}, 924--933.

\bibitem[{Solovieff et~al.(2013)Solovieff, Cotsapas, Lee, Purcell \&
  Smoller}]{solovieff_pleiotropy_2013}
\textsc{Solovieff, N.}, \textsc{Cotsapas, C.}, \textsc{Lee, P.~H.},
  \textsc{Purcell, S.~M.} \& \textsc{Smoller, J.~W.} (2013).
\newblock Pleiotropy in complex traits: challenges and strategies.
\newblock \textit{Nature Reviews Genetics} \textbf{14}, 483--495.

\bibitem[{Staiger \& Stock(1997)}]{staiger_instrumental_1997}
\textsc{Staiger, D.} \& \textsc{Stock, J.~H.} (1997).
\newblock Instrumental variables regression with weak instruments.
\newblock \textit{Econometrica} \textbf{65}, 557--586.

\bibitem[{Stiglitz(1976)}]{stiglitz_efficiency_1976}
\textsc{Stiglitz, J.~E.} (1976).
\newblock The efficiency wage hypothesis, surplus labour, and the distribution
  of income in ldcs.
\newblock \textit{Oxford economic papers} \textbf{28}, 185--207.

\bibitem[{Stock et~al.(2002)Stock, Wright \& Yogo}]{stock_survey_2002}
\textsc{Stock, J.~H.}, \textsc{Wright, J.~H.} \& \textsc{Yogo, M.} (2002).
\newblock A survey of weak instruments and weak identification in generalized
  method of moments.
\newblock \textit{Journal of Business \& Economic Statistics} \textbf{20}.

\bibitem[{Subramanian \& Deaton(1996)}]{subramanian_demand_1996}
\textsc{Subramanian, S.} \& \textsc{Deaton, A.} (1996).
\newblock The demand for food and calories.
\newblock \textit{Journal of political economy} , 133--162.

\bibitem[{Tan(2006)}]{tan_regression_2006}
\textsc{Tan, Z.} (2006).
\newblock Regression and weighting methods for causal inference using
  instrumental variables.
\newblock \textit{Journal of the American Statistical Association}
  \textbf{101}, 1607--1618.

\bibitem[{Tiffin \& Dawson(2002)}]{tiffin_demand_2002}
\textsc{Tiffin, R.} \& \textsc{Dawson, P.~J.} (2002).
\newblock The demand for calories: some further estimates from zimbabwe.
\newblock \textit{Journal of Agricultural Economics} \textbf{53}, 221--232.

\bibitem[{Wooldridge(2010)}]{wooldridge_econometrics_2010}
\textsc{Wooldridge, J.~M.} (2010).
\newblock \textit{Econometric Analysis of Cross Section and Panel Data}.
\newblock MIT press, 2nd ed.

\end{thebibliography}

\end{document}